# Using a Multi-Cathode Counter (MCC) in the Search for Hidden Photon CDM.

A.V.Kopylov, I.V.Orekhov, and V.V.Petukhov

*Institute for Nuclear Research of Russian Academy of Sciences,*

*117312, Prospect of 60$^{th}$ October Revolution 7A*

*Moscow, Russia*

We report on a new technique of a Multi-Cathode Counter (MCC) developed to search for hidden photon (HP) cold dark matter (CDM) with a mass from 5 to 10 eV. The method suggested in the assumption that hidden photons of the mass greater than a work function of the metal, the cathode of the counter is fabricated induce emission of single electrons from a cathode. Three configurations of the same counter are used to measure the count rates $R_1$, $R_2$ and $R_3$ of the single electron events sequentially each week, where the difference $R_1 - R_2$ measures the effect from HP and $R_3$ is used as a reference sequence to monitor the counting process. At present the work is aimed to refine the procedure of data treatment and to look for long term variations which should be accounted for in the final analysis of data.

## 1. Introduction.

The present observational cosmology revealed that our visible world is "immersed" in a non baryonic matter (dark matter, DM) with mass higher than a mass of the visible world but we don't know what would be the nature of this substance. In other words we can't say what would be the matter our universe is made of. This certainly is absolutely inappropriate situation and presents probably the main challenge for experiment to-day. Theoreticians are suggesting a number of candidates for DM among them WIMPs which should produce nuclear recoils in matter and the experiments on WIMPs are in the main stream of the search. However, there are other candidates, among them axion dark matter or HP (hidden photon, $\tilde{X}^\mu$) CDM which is light extra gauge boson. It may be observed in experiment through a kinetic mixing $(\chi/2)F_{\mu\nu}\tilde{X}^{\mu\nu}$ with the ordinary photons. It has been shown that there is still a huge region of allowed parameters for hidden photon; see for example Ref [1] and references therein. A number of experiments are searching for these objects, for example, axions are searched in the Axion Dark Matter eXperiment (ADMX) by means of a resonant cavity and magnetic field [2]. Recently the eV mass range of HP was investigated with a dish antenna [3], a novel method proposed in Ref. [4]. The idea is to detect the reflected from metallic mirror (antenna) electromagnetic wave which is emitted by the oscillation of electrons of the antenna's surface under the tiny electric field of the HP. This method works well only if the reflectance of mirror is high. Here in our work we focused on shorter wavelengths, i.e. higher masses of HP for which the reflectance of mirror is low. It is assumed that in this case the oscillation of electrons of the antenna's surface under the tiny electric field of HP will induce with certain quantum efficiency η the emission of single electrons from a mirror. This probability is taken to be equal to the one of the emission of single electrons from a metal by real photons with energy ω = $m_{\gamma'}$. We have developed a special technique of Multi-Cathode Counter (MCC) for recording of this kind of events and subtracting of a background.

## 2. Experimental apparatus.

The general view of the counter is presented on Fig.1 and the electronic scheme on Fig.2. The cathode of the counter is 194 mm in diameter and 400 mm in length. It has relatively large (≈ 0.1 $m^2$) surface which acts in this experiment as a "mirror" for HP but instead of reflecting light it emits single electrons. It has a central anode wire of 20 μm and 4 cathodes, 3 of them are composed of an array of 100 μm nichrome wires tensed with a pitch of a few mm around the anode wire one after another, and a fourth one, more distant from anode, is a cathode, a mirror, which is made of copper in the form of a cylinder.

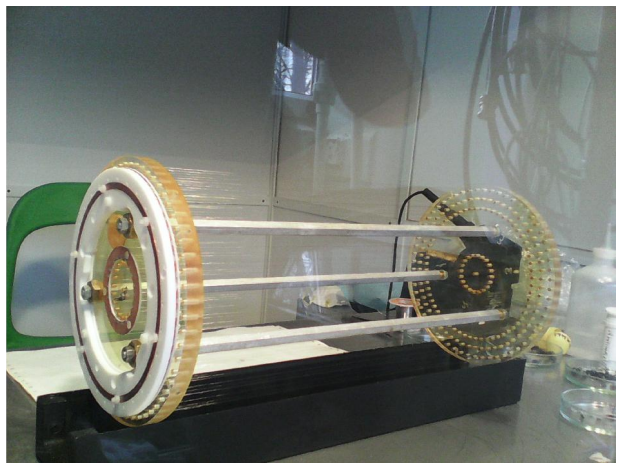

Figure 1. The general view of the counter.

The apparatus is counting electrons emitted from the walls of a "mirror" at short wavelengths ω = $m_{\gamma'}$ ≈ 10 eV when the reflectance of a "mirror" is low. The counter is filled by gas, argon at 0.2 MPa in this particular work. The diameter of the first cathode is 40 mm to ensure high ($\approx 10^5$) coefficient of gas amplification. Third cathode of the diameter 180 mm is used for measurements of the background and second cathode of 140 mm in diameter is used to collect the independent series of data which serves as a "reference" data and enables to put under control the variations due to some changing conditions of counting which may affect the count rates. The counter is used in three different configurations. In the first configuration 4$^{th}$ cathode is under the highest negative voltage to ensure that all electrons emitted from the copper surface would drift to the central section with high gas amplification. This configuration is used to measure the rate $R_1$ of single electrons emitted from a copper cathode. In the second configuration the highest negative voltage is applied to the 3$^{rd}$ cathode of diameter 180 mm. In this configuration 4$^{th}$ cathode, a copper cylinder, is grounded so that electrons of low energy, less then a few keV, emitted from the copper are scattered back in argon at 0.2 MPa by high voltage applied to the 3$^{rd}$ cathode. The counter in this configuration is used to measure the background count rate $R_2$. So, opposite to a common practice of using a central part of the detecting volume for measurement of the effect, here only a small layer adjacent to the copper cathode is used to measure the effect, the rest of the sensitive volume serves for the measurement of the background. It enables to subtract from the measured effect the contribution from the ends of the counter where electric fields are distorted and also the one from emission of single electrons from multiple nichrome wires. In the third configuration the highest negative voltage is applied to 2$^{nd}$ cathode and a copper cylinder is grounded. This configuration is used to provide the independent series of data $R_3$ to monitor the external conditions of counting which may affect the counting process.

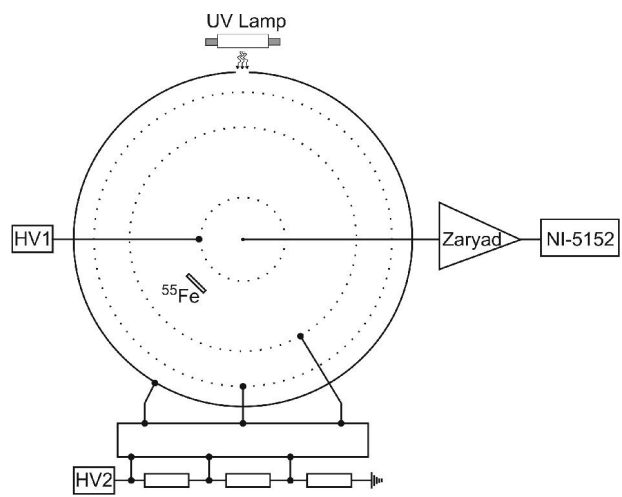

Figure 2. The electronic scheme.

The measurements of $R_1$, $R_2$ and $R_3$ are performed sequentially one after another and the average is drawn for each value for each week of measurements. We assume that during a week there are no changes in count rates. The difference $R_1$-$R_2$ gives the value of the effect measured for this particular week. The design of the counter should be so optimized that it enables for the difference to be close to zero if there is no effect from HP. In reality there will be always some difference which is just unaccountable background. But because we use a difference of two measured count rates for the same counter in the same lay-out some systematic uncertainties in measurement of $R_1$ and $R_2$ are partially cancelled because the geometry of the configurations 1 and 2 are very similar. In the measurements the shapes of the pulses on the output of a charge sensitive preamplifier are recorded by 8-bit digitizer. Pulse shape discrimination is used to select "true" pulses from "noise" pulses similar to what has been done in [5, 6]. To reduce the background from external γ-radiation the counter is placed in a steel cabinet with 30 cm iron shield placed at a ground floor of a building in Troitsk, Moscow.

## 3. Energy calibration and analysis.

The calibration of the counter has been conducted by $^{55}$Fe source and by UV light of the mercury lamp. The source has been placed inside the counter between first and second cathodes facing the anode wire. High voltage at first cathode was 2060 V and high voltages applied to voltage divider has been used for all three configurations such as to ensure the amplitude of the pulse corresponding to 5.9 keV of $^{55}$Fe on the output of charge sensitive preamplifier to be at the level 1600 mV what corresponds to a gas amplification of about $10^5$. Then the source has been removed from the counter and the internal walls of the counter were irradiated by UV light from a mercury lamp placed outside through a window made of melted silica. Figure 3 shows the single electron spectra obtained in measurements in all 3 configurations. Comparing the count rates $R_1$ and $R_2$ one can see that in the 2$^{nd}$ configuration the count rate of single electrons was about 10 times lower than in the first one. It proves that in the 2$^{nd}$ configuration the electrons emitted from the walls of the counter were really rejected back by the 3$^{rd}$ cathode. It means that the counter in the 2$^{nd}$ configuration can be used for measurement of the background of the single electron count rates measured in the first configuration. The data were collected frame by frame. Each frame contained 2M points each point 100 ns. After collection of the data they were stored on a disk then the collection resumed. The analysis of the collected data is performed off-line. It consists in searching for small peaks of the "true" shape with the amplitude from 3 to 30 mV which correspond to energies from 12 to 120 eV. The frames with the signs of excessive noisiness were removed from analysis.

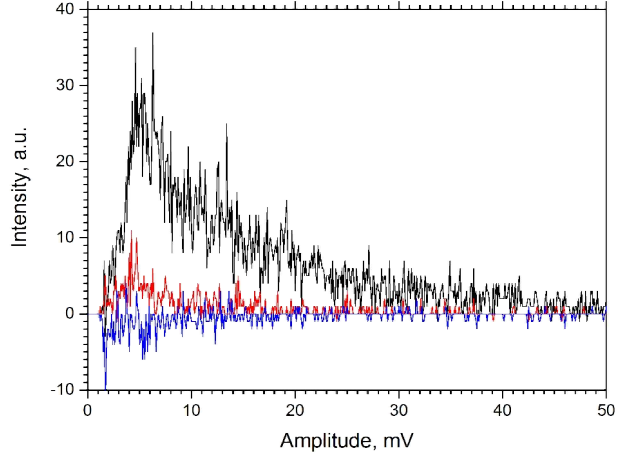

Figure 3. Single-electron spectra from UV of mercury lamp obtained for 3 different configurations D1 (black), D2 (red) and D3 (blue).

## 4. Sensitivity of the method.

The experiment is conducted in the assumption that hidden photons of the mass greater than a work function of a metal, the cathode of the counter is fabricated induce the oscillation of the current on the surface of a cathode which induces emission of electrons from the surface of a cathode with some probability η. If we take this probability η equal to the probability for the real photon of energy $E_\gamma = m_{\gamma'} \approx 10$ eV to eject an electron, then we can easily transform the sensitivity from the case of a dish antenna to our case of the counter with a cathode as a "mirror". Taking the sensitivity from Ref. [4], this is written here for the case when all dark matter is composed of hidden photons,

$$\chi_{sens} = 5.6 \cdot 10^{-12} \left(\frac{R_{\gamma,\det}}{1\ Hz}\right)^{\frac{1}{2}} \left(\frac{m_{\gamma'}}{eV}\right)^{\frac{1}{2}} \left(\frac{0.3\ GeV/cm^3}{\rho_{CDM,halo}}\right)^{\frac{1}{2}} \left(\frac{1\ m^2}{A_{dish}}\right)^{\frac{1}{2}} \left(\frac{\sqrt{2/3}}{\alpha}\right)$$

where $R_{\gamma,\det}$ is the minimum count rate from HP which can be observed in the experiment one can see that for $m_{\gamma'} = 10$ eV and $A_{dish} = 0.1$ m$^2$ to set a limit on the level of $10^{-11}$ one should have $R_{\gamma,\det}$ on the level of 0.1 Hz. The difference $R_1$-$R_2$ should be in this case lower than this value by a factor 1/η. In case of a dish antenna η is a quantum efficiency of a photocathode of PMT and according to Ref [2] PMT they used had a sensitive range 300-650 nm and a peak quantum efficiency of 17%. In our case at the energies $E_\gamma = m_{\gamma'} \approx 10$ eV which correspond to wavelength 120 nm the probability that electron will be ejected η ≈ 1%. Thus the difference $R_1$-$R_2$ should be in this case on the level of $10^{-3}$ Hz. The sensitivity depends on η as a square root but certainly, in comparison with a dish antenna we somewhat loose in sensitivity due to lower quantum efficiency η. We propose here a complimentary method: in comparison with a dish antenna we can cover adjacent region of higher masses for HP and if we use for the cathode of the counter several materials with different work functions principally it will be possible to look into each narrow interval of the mass range from 5 to 10 eV.

## 5. Conclusion.

A new technique of Multi Cathode Counter (MCC) has been developed to search for hidden photon CDM in the assumption that all dark matter is composed of hidden photons (HP). It was assumed also that HP of the mass greater than a work function of the metal, the cathode of the counter is fabricated induce emission of single electrons from a cathode. The technique is planned to be used to search for HP with a mass from 5 to 10 eV. At present the work is aimed to refine the procedure of data treatment and to look for long term variations which should be accounted for in the final analysis of data.

## Acknowledgements.

The authors express deep gratitude to E.P.Petrov for continuous help during this work, to A.I.Egorov for skill and patience during fabrication of many different versions of the counter and to Grant of Russian Government "Leading Scientific Schools of Russia" #3110.2014.2 for partial support of this work.

## References.

1. Michael Klasen, Martin Pohl, Gunter Sigl / Indirect and direct search for dark matter // arXiv:1507.03800
2. H.Peng, S.Asztalos, E.Daw, N.A.Golubev, C.Hagman et al. // NIM A 444 (1999) 569
3. J.Suzuki, T.Horie, Y.Inoue, and M.Minowa / Experimental Search for Hidden Photon CDM in the eV mass range with a Dish Antenna // arXiv:1504.00118v1 [hep-ex]
4. D. Horns, J. Jaeckel, A. Lindner, A. Lobanov, J.Redondo, and A.Ringwald / Searching for WISPy Cold Dark Matter with a Dish Antenna // JCAP 04 (2013) 016: arXiv: 1212.2970
5. A.V.Kopylov, I.V.Orekhov, V.V.Petukhov, and A.E.Solomatin / Gaseous detector of ionizing radiation for registration of coherent scattering on nuclei // Technical Physics Letters, Volume: 40 Issue: 3 Pages: 185-187 (2014); arXiv: 1311.6564

6. A.V.Kopylov, I.V.Orekhov, V.V.Petukhov, and A.E.Solomatin / Gaseous Detector with sub-keV Threshold to Study Neutrino Scattering at Low Recoil Energies // Advances of High Energy Physics 147046 (2014); arXiv: 1409.4873